\documentclass[10pt,aps,prd,nofootinbib,superscriptaddress,twocolumn]{revtex4}
\usepackage{amssymb,amsmath,latexsym,graphics, graphicx,epsfig,multirow,comment,hyperref,appendix} 

\newcommand{\gsim}{\lower.7ex\hbox{$\;\stackrel{\textstyle>}{\sim}\;$}}
\newcommand{\lsim}{\lower.7ex\hbox{$\;\stackrel{\textstyle<}{\sim}\;$}}

\def\OO{{\cal O}}

\newcommand{\keV}{\,\mathrm{keV}}

\newcommand{\bef}{\begin{figure}[htbp]\begin{center}}
\newcommand{\eef}{\end{center}\end{figure}}

\newcommand{\dm}{{\text{dm}}}

\begin{document}

\pagestyle{plain}

\title{
\begin{flushright}
\mbox{\normalsize SLAC-PUB-13802}
\end{flushright}
\vskip 15 pt

Disentangling Dark Matter Dynamics with Directional Detection}

\author{Mariangela Lisanti and Jay G. Wacker }

\affiliation{
Theory Group, SLAC,  Menlo Park, CA 94025
}

\begin{abstract}
Inelastic dark matter reconciles the DAMA anomaly with other null direct detection experiments and
points to a non-minimal structure in the dark matter sector.
In addition to the dominant inelastic interaction, dark matter scattering may have a subdominant elastic component.  If these elastic interactions are suppressed at low momentum transfer, they will have similar nuclear recoil spectra to inelastic scattering events.  While upcoming direct detection experiments will see strong signals from such models, they may not be able to unambiguously determine the presence of the subdominant elastic scattering from the recoil spectra alone.  We show that directional detection experiments can separate elastic and inelastic scattering events and discover the underlying dynamics of dark matter models.
\end{abstract}
\pacs{} \maketitle

\section{Introduction}

The annual modulation anomaly from the DAMA experiment is an intriguing hint of the identity of dark matter \cite{Bernabei:2005hj,Bernabei:2008yi}.  Direct detection experiments such as DAMA measure the energy of nuclear recoils from incident dark matter particles.  The signal experiences an annual modulation because of the variation of the Earth's relative motion with respect to the dark matter in the halo.  While the DAMA experiment has measured an $8.2\sigma$ modulation with the correct phase, its results are in conflict with those from  CDMS  \cite{Akerib:2005za, Akerib:2005kh, Ahmed:2008eu}, XENON10 \cite{XENON10, Collaboration:2009xb}, CRESST \cite{Angloher:2004tr, CRESSTII}, ZEPLIN-II \cite{Alner:2007ja}, and ZEPLIN-III \cite{ZEPLINIII}, which measure the total, unmodulated scattering rate.

All current direct detection experiments are optimized to look for elastic scattering, which has an exponentially falling recoil energy spectrum \cite{Lewin:1995rx}.  A distinguishing feature of DAMA's measured modulation spectrum is that it is suppressed at energies below $\sim$ 25 keVnr, where elastic events should dominate.  DAMA's results, taken together with the null results from all other direct detection experiments, can be  elegantly explained if the dark matter scatters inelastically off of nuclei to an adjacent state that is $\OO(100 \keV)$ more massive than the ground state \cite{TuckerSmith:2001hy,Chang:2008gd, Cui:2009xq,SchmidtHoberg:2009}.  In the case of inelastic dark matter (iDM), a minimum velocity is needed to upscatter and the recoil spectrum is suppressed below this kinematic threshold.  

Inelastic dark matter has three features that lead to the consistency of all current experiments.  First, iDM has a much larger dependence on the Earth's velocity than elastic dark matter and thus the annual modulation is often 25\% or larger, in comparison to $2.5\%$ for elastic scattering.  This decreases DAMA's unmodulated cross section
by a factor of 10 or more and therefore reduces the expected signal by the same factor in all experiments looking for unmodulated scattering.
Second, iDM dominantly scatters off heavier nuclei and lighter nuclei may not have enough kinetic energy to excite the
transition.  As a result, CDMS' sensitivity is strongly suppressed due to inelastic kinematics.  Lastly, inelastic scattering events have higher recoil energies than elastic events.  Recent experiments such as XENON10 and ZEPLIN-III have shrunk their recoil energy window to eliminate higher energy scattering events, reducing the acceptance for iDM.  
However, XENON10 has recently reanalyzed their data over a larger signal window to be sensitive to iDM and their latest results are used in this article \cite{Collaboration:2009xb}. 


A dynamical alternative to the inelastic hypothesis was recently discussed in \cite{Chang:2009yt,Feldstein:2009tr}; in this scenario, the interaction between the dark matter and Standard Model (SM) is elastic, but is suppressed by a form factor 
\begin{equation}
F_{\dm}(q^2) = c_0 + c_1 q^2 + c_2 q^4 + \cdots.
\end{equation}
If the first term in this expansion vanishes, the elastic scattering rate is multiplied by additional factors of $q^2 = 2 m_N E_R$ and goes to zero as $E_R \rightarrow 0$.  Form factor-suppressed scattering can arise from multipole, polarization, and charge-radius interactions between the dark sector and the Standard Model \cite{Pospelov:2000bq}.  The recoil spectrum for form factor elastic dark matter (FFeDM) is suppressed at low energies and more closely resembles the spectrum for inelastic events rather than standard elastic events, which peaks at low energies.  However, it is more challenging to reconcile DAMA and the null experiments using only FFeDM because the form factor depends on the product $m_N E_R$ and any suppression for light nuclei can be compensated by looking at larger energies.  

\begin{figure}[b] 
   \centering
   \includegraphics[width=3.4in]{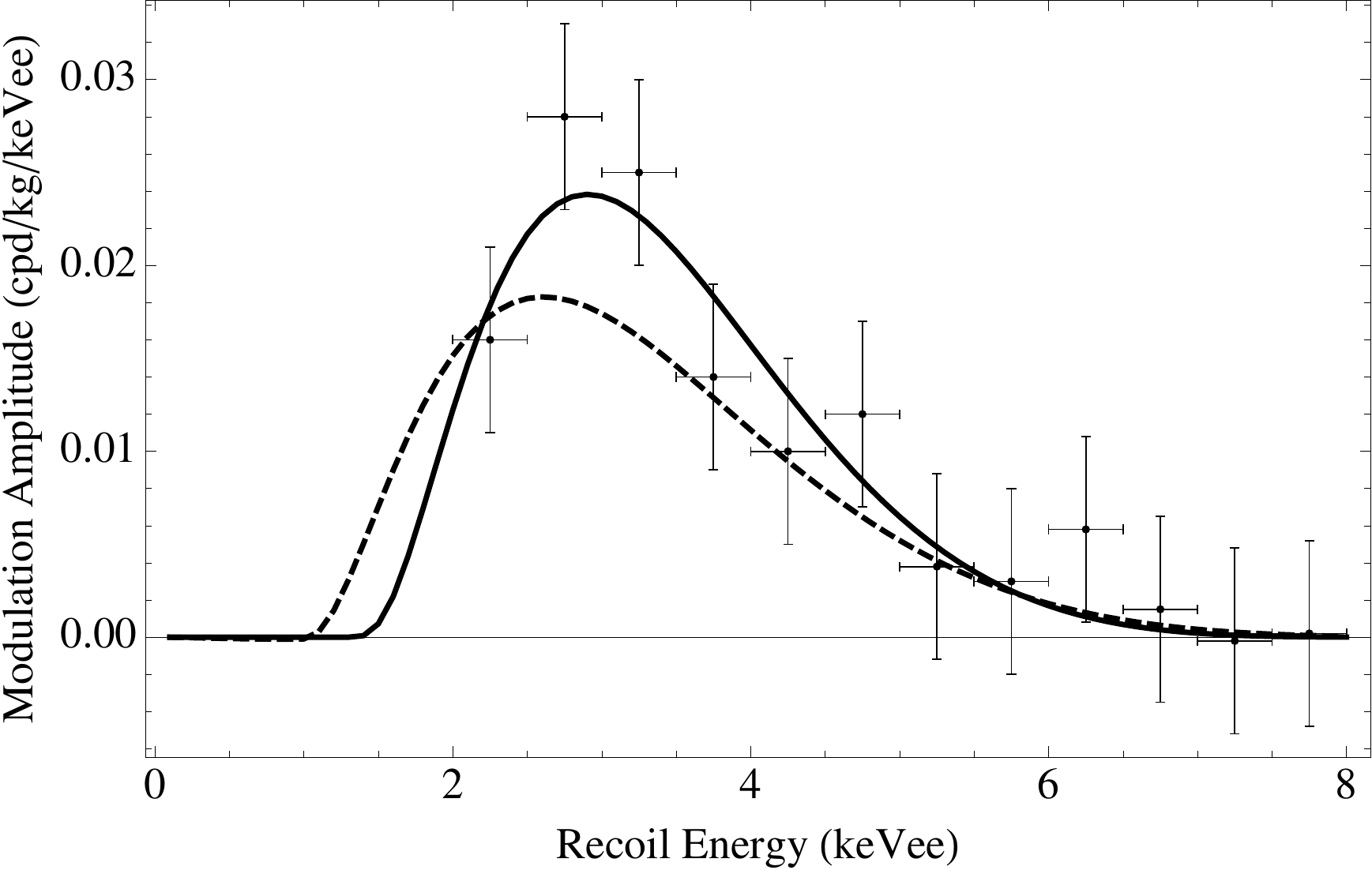} 
   \caption{Comparison of the modulation amplitude for two scattering scenarios: completely inelastic scattering (solid) and inelastic scattering with a subdominant elastic charge-radius component (dashed).  The points show the modulation amplitude measured by DAMA and DAMA/LIBRA \cite{Bernabei:2005hj}.  The electron equivalent energy (keVee) is the recoil energy rescaled by the quenching factor for iodine ($q_I = 0.085$). } 
\label{Fig: AnnMod}
\end{figure}

In this paper, we consider a scenario where inelastic scattering is complemented by a form factor-suppressed elastic component.  
These scenarios arise, for example, in composite dark matter models \cite{Lisanti:2009} and yield new challenges for direct detection phenomenology. As is illustrated in Fig.~\ref{Fig: AnnMod}, the modulation amplitude for inelastic scattering with subdominant charge-radius scattering (dashed 
line) resembles that for inelastic scattering (solid line) and is markedly different from the regular elastic spectrum.  Distinguishing the elastic and inelastic contributions is critical for understanding the underlying structure of the theory.  This article addresses how directional detection experiments, which can measure the direction of the nuclear recoil in addition to its energy \cite{Sciolla:2008vp,Finkbeiner:2009ug}, can differentiate the contributions to the scattering rate.  In the following section, we review the direct detection phenomenology.  In Sect.~\ref{Sec: Directional Detection}, we introduce directional detection experiments and show how they can distinguish the dynamics in the dark sector.  We conclude in Sec.~\ref{Sec: Discussion}.

\section{Direct Detection Phenomenology}
\label{Sec: Direct Detection}

Direct detection experiments measure the energy of nuclear recoils in a detector.  For a detector consisting of nuclei with mass $m_N$, the differential scattering rate per unit detector mass is 
\begin{eqnarray}
\frac{dR}{dE_R}=\frac{\rho_0}{m_{\dm} m_N}\left\langle \frac{d \sigma}{d E_R} v \right \rangle,
\label{eq: rate}
\end{eqnarray}
where $\rho_0 = 0.3$ GeV/cm$^3$ is the local dark matter density.  The rate depends on the dark matter mass $m_{\text{dm}}$, as well as the differential cross section.  When both elastic and inelastic scattering are allowed, the differential cross section is parameterized by
\begin{eqnarray}
\frac{d\sigma}{dE_R} &=&  \left[ c_{\text{el}}^2 \Big(\frac{2 m_{\text{N}}E_R}{\Lambda^{2}}\Big)^{n_{\text{el}}}  +  c_{\text{in}}^2 \Big(\frac{2 m_{\text{N}}E_R}{\Lambda^2} \Big)^{n_{\text{in}}}\right]
\frac{d\sigma_0}{d E_R}, \nonumber \\
\label{Eq: Sigma}
\end{eqnarray}
where $c_{\text{el,in}}$ are the dimensionless couplings for the elastic and inelastic interactions, $\Lambda$ is the scale for new physics, and 
\begin{eqnarray}
\frac{d \sigma_0}{d E_R} = \frac{m_N}{2v^2} \frac{\sigma_\text{N}}{\mu^2} \frac{(f_\text{p} Z + f_\text{n}(A-Z))^2}{f_\text{p}^2}
|F_{\text{H}}(E_R)|^2
\label{eq: xsection}
\end{eqnarray}
is the standard rate for elastic scattering off a nucleus with charge $Z$ and atomic number $A$.  Here, $\mu$ is the reduced mass of the dark matter-nucleus system,  $\sigma_{\text{N}}$ is the cross section for the dark matter-nucleus interaction at zero momentum transfer, and $f_{\text{p,n}}$ are the couplings to the proton and neutron, respectively.  Our results are normalized to $f_\text{p} = 1$ and $f_\text{n} = 0$, and we assume that $\Lambda = 1$ GeV.  

The scattering operators coherently couple the dark matter states to the nuclear charge, and the Helm form factor accounts for loss of coherence at large momentum transfer, $q$,
\begin{eqnarray}
|F_{\text{H}}(E_R)|^2 = \left(\frac{3j_1(|q|r_0)}{|q|r_0}\right)^2 e^{-s^2|q|^2},
\end{eqnarray}
where $s=1$ fm, $r_0=\sqrt{r^2-5s^2}$, and $r=1.2A^{1/3}$ fm  \cite{Helm:1956zz}.

The specific values of $n_{\text{in}}$ and $n_{\text{el}}$ are model-dependent.  For instance, regular iDM has $n_{\text{in}} = 0$, while regular elastic scattering corresponds to $n_{\text{el}} = 0$.  Form factor suppression in either the inelastic or elastic scattering interactions results in additional powers of $E_R$.   Composite inelastic dark matter models have $n_{\text{in}}=1$, though this does not make a substantive change to the spectrum as long as $ \Lambda^2 \gg m_N E_R$ \cite{Alves:2009nf,Kaplan:2009de}. 
In this paper, we focus on the scenario where inelastic scattering ($n_{\text{in}} = 1$) is supplemented by a form factor-suppressed elastic component ($n_{\text{el}}=2$), although all the conclusions are general and apply to $n_{\text{in}}=0$. 

For a given dark matter velocity distribution function $f(v)$ defined in the galactic rest frame,
\begin{eqnarray}
\left\langle \frac{d \sigma}{d E_R} v \right \rangle =\int_{v_{\text{min}}}\!\!\!d^3v\; f(\vec{v} + \vec{v}_e)\; v\frac{d\sigma}{dE_R}.
\label{eqn: vavg}
\end{eqnarray}
The minimum velocity is set by the kinematics of the scattering process:
\begin{eqnarray}
v_\text{min}(E_R) = \begin{cases}
\sqrt{\frac{m_N E_R}{2 \mu^2}} & \text{elastic}\\
\frac{1}{\sqrt{2 m_N E_R}} \Big(\frac{m_N E_R}{\mu} + \delta m\Big), & \text{inelastic}
\end{cases}
\end{eqnarray}
where $\delta m$ is the dark matter mass splitting.
The Earth's velocity in the galactic rest frame, $\vec{v}_e$, is defined as 
\begin{eqnarray}
\vec{v}_e = \vec{v}_{\odot} + \vec{v}_{\oplus}(t).
\end{eqnarray}
In the coordinate system where $\hat{x}$ points towards the galactic center, $\hat{y}$ points in the direction of the galactic rotation, and $\hat{z}$ points towards the galactic north pole, the sum of the Sun's local Keplerian velocity and its peculiar velocity is \cite{Dehnen:1997cq, Binney}
\begin{eqnarray}
\vec{v}_\odot(t) \approx  (0, 220, 0) + ( 10, 5, 7)  \text{ km/s}.
\end{eqnarray}
The velocity of the Earth in the sun's rest frame is given by
 \begin{eqnarray}
\vec{v}_\oplus \approx v_{\oplus} \left[\hat{\epsilon}_1 \cos \frac{2 \pi (t - t_0)}{\text{yr}} +  \hat{\epsilon}_2 \sin \frac{2 \pi (t - t_0)}{\text{yr}}  \right],
 \end{eqnarray}
where $t_0$ is the spring equinox ($\approx$ March 21), $v_{\oplus} = 29.8$ km/s is the orbital speed of the Earth \cite{Lewin:1995rx}, and
\begin{eqnarray}
\hat{\epsilon}_1 &=& (0.9931, 0.1170, -0.01032)  \\ \nonumber
\hat{\epsilon}_2 &=& (-0.0670,0.4927,-0.8676) 
\end{eqnarray}
are the unit vectors in the direction of the Sun at the spring equinox and summer solstice, respectively \cite{Gelmini:2000dm, Savage:2008er}.  

Typically, the velocity distribution function $f(v)$ is assumed to be isothermal and isotropic in the galactic frame.  In the ``Standard Halo Model'' (SHM), the Keplerian velocity is constant throughout the galaxy and the velocity dispersion is assumed to be Gaussian.  These assumptions are not consistent with recent N-body simulations of dark matter particles; simulations such as Via Lactea \cite{Diemand:2006ik} show that the dark matter velocity dispersion is anisotropic and that the density falls off more steeply at larger radii than in the isothermal scenario.  

This article adopts the following {\it ansatz} for the dark matter halo velocity distribution
\begin{equation}
f(v)= N \Big(e^{ - (v/ v_0)^{2\alpha}} - e^{- (v_{\text{esc}}/v_0)^{2\alpha}} \Big) \;\Theta(v_{\text{esc}}-v),
\label{eq: halo}
\end{equation}
which reproduces the SHM in the limit $\alpha \rightarrow 1$.  This distribution function captures the most important qualitative features of Via Lactea, but in an analytical form that is easy to compute with. The parameter $\alpha$ changes the shape of the profile near the escape velocity.  Inelastic scattering events are particularly sensitive to the high-velocity tail because their minimum velocity is larger in comparison to the elastic case \cite{MarchRussell:2008dy, Fairbairn:2008gz}.   

To find the regions of parameter space consistent with current direct detection experiments, we perform a global chi-squared analysis and marginalize over the six unknown parameters: $v_0, v_{\text{esc}}, \alpha, m_{\dm}, \delta m,$ and $\sigma_{p}$, the cross section per nucleon.  The $\chi^2$ function is 
\begin{equation}
\chi^2 (m_{\dm}, \delta m, \sigma_n, v_0, v_{\text{esc}}, \alpha) = \sum_{i=1}^{N_{\text{exp}}} \Bigg( \frac{X_i^{\text{pred}} - X_i^{\text{obs}}}{\sigma_i} \Bigg),
\label{eq: chi2}
\end{equation}
where $N_{\text{exp}}$ is the number of experiments included in the fit, $X_i^{\text{pred}}$ is the predicted experimental result, $X_i^{\text{obs}}$ is the observed result and $\sigma_i$ is the known error.  The velocity distribution parameters are constrained to be within
\begin{eqnarray}
\nonumber
200 \text{ km/s} \le v_0 \le 300 \text{ km/s}\\
\nonumber
500 \text{ km/s} \le v_{\text{esc}} \le 600 \text{ km/s}\\
0.8 \le \alpha \le 1.25.
\end{eqnarray}
These values are motivated by rough observational constraints  \cite{Smith:2006ym, Lewin:1995rx} and the fact that (\ref{eq: halo}) is a spherically symmetric form of the Via Lactea fits by \cite{Fairbairn:2008gz}.   

The results are fit to the first twelve bins (2-8 keVee) of the recoil spectrum measured by DAMA \cite{Bernabei:2005hj}.  The modulation  amplitude from 8-14 keVee is combined into a single bin with amplitude $-0.0002\pm.0014$ cpd/kg/keVee.  
The contribution of this last bin to the total $\chi^2$ is typically negligible.  The region of parameter space that is consistent with DAMA is constrained by the null experiments, each of which has observed a number of events in its signal window.  The results from  CDMS, CRESST, ZEPLIN-II, and ZEPLIN-III are included in the fit, in addition to XENON10's recently updated analysis with an expanded signal window \cite{Collaboration:2009xb}.  We require that the theory not saturate the number of observed events for each experiment to 95\% confidence.  

\begin{figure}[tb] 
   \centering
   \includegraphics[width=3.4in]{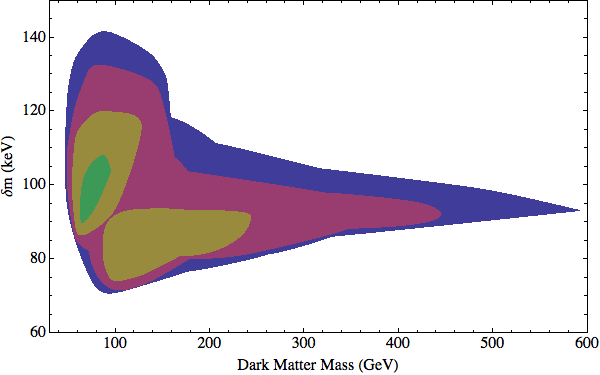} 
   \caption{The 95\% contours in the $m_{\dm} - \delta m$ parameter space corresponding to $c_{\text{el}}/c_{\text{in}}$ = 0 (blue), 0.4 (magenta), 0.6 (yellow) and 0.8 (green) for the scenario with $n_{\text{in}} = 1$ and $n_{\text{el}}=2$.  The four regions overlap one another.  }
   \label{fig: parspace}
\end{figure}

Figure~\ref{fig: parspace} shows the allowed regions of $m_{\text{dm}} - \delta m$ parameter space for different ratios of elastic to inelastic scattering.  The contours are defined as 
\begin{equation}
\chi^2 (m_{\dm}, \delta m, \sigma_n, v_0, v_{\text{esc}}, \alpha) =  \chi^2_{\text{min}} + \Delta \chi^2(\text{CL}),
\end{equation}
where $\Delta \chi^2(95\%) = 12.6$ for six degrees of freedom.  The minimal $\chi^2$ found was $\chi_{\text{min}}^2 = 4.1$ for $c_{\text{el}}/ c_{\text{in}}$= 0 and thus every prediction was forced to have a $\chi^2 \le 16.7$ corresponding to the 2$\sigma$ band.  

The best-fit point for the completely inelastic scenario is
\begin{eqnarray}
\label{Eq: BestFit iDM}
&&( v_0, v_{\text{esc}}, \alpha)  =  (279, 586, 0.80)   \\
&& (m_{\dm} , \delta m, \sigma_n )= ( 59 \text{ GeV}, 121 \text{ keV}, 5.9 \times 10^{-39} \text{ cm}^2), \nonumber 
\end{eqnarray}
where $\sigma_n=\sigma_\text{N} (\mu_N^2/\mu_n^2)$ and $\mu_{\text{N}(n)}$ is the reduced mass of the dark matter-nucleus (nucleon) system.  As the ratio of elastic to inelastic scattering increases, the null experiments become more constraining and the minimum $\chi^2$ increases.  The maximum allowed ratio of elastic to inelastic scattering is $c_{\text{el}}/c_{\text{in}} = 0.8$.  For this ratio, the best-fit point has $\chi^2 = 14.3$ and corresponds to
\begin{eqnarray}
\label{Eq: BestFit iDM FFeDM}
&( v_0, v_{\text{esc}}, \alpha)  =  (297, 500, 0.82) &  \\ 
& (m_{\dm} , \delta m, \sigma_n )= ( 74 \text{ GeV}, 98 \text{ keV}, 1.2 \times 10^{-39} \text{ cm}^2).&\nonumber
\end{eqnarray}
There is a high degree of degeneracy in the allowed values of the halo parameters for each $m_{\text{dm}} - \delta m$ point.  
The parameter points with the least tension have $m_{\text{dm}} $ and $\delta m$ near the best-fit points of (\ref{Eq: BestFit iDM}) and (\ref{Eq: BestFit iDM FFeDM}), but the halo parameters and cross sections can vary wildly.   Some areas of the allowed regions are highly sensitive to 
the exact range of the halo parameters.  For instance, the large mass region corresponds to exceedingly ``slow'' velocity
distribution functions where $v_0 \lsim 225 \text{ km/s}$ and $ \alpha \gsim 1.15$.

\begin{figure}[tb] 
   \centering
   \includegraphics[width=3.4in]{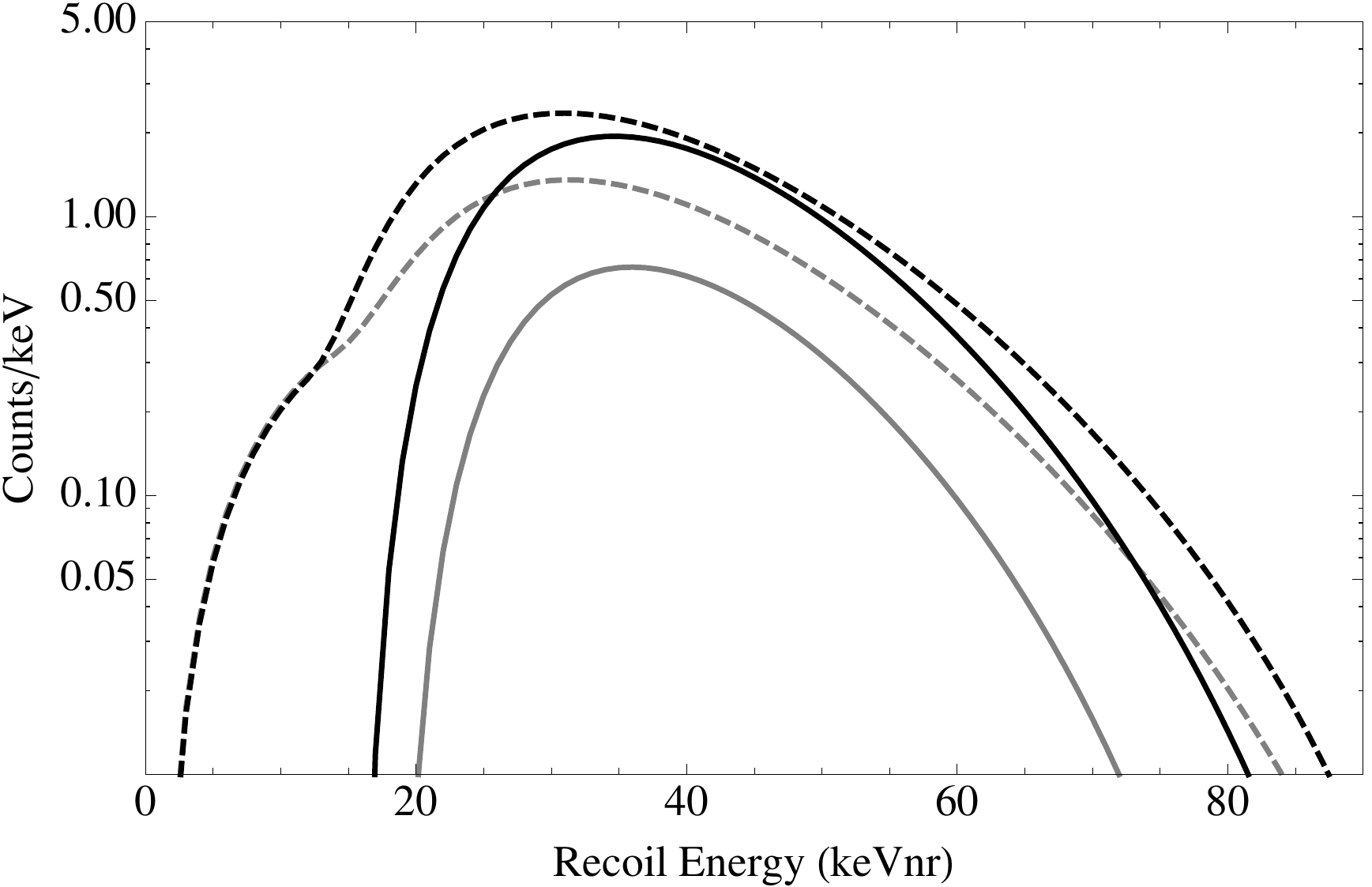}
   \caption{Estimated recoil spectrum at LUX with 1000 kg-days and $30\%$ efficiency during the summer (black) and winter (gray).  The solid line corresponds to the best-fit point for completely inelastic scattering; the dashed line corresponds to the best-fit point when $ c_{\text{el}}= 0.8 c_{\text{in}}$. }
   \label{Fig: Example}
\end{figure}

In the near future, definitive experiments such as XENON100 \cite{Xenon100} and LUX \cite{LUX} will confirm or refute the inelastic dark matter hypothesis.  These two experiments are the upgrades to the XENON10 detector.  The LUX detector will consist of 300 kg of liquid xenon with a planned fiducial volume of 100 kg, and it will be sensitive to recoil energies as large as $\sim$ 300 keVnr \cite{Sorenson}.  Figure~\ref{Fig: Example} shows an estimated recoil spectrum at LUX after 1000 kg-days and assuming 30\% efficiency.  The solid line denotes the spectrum for the best-fit point in the pure inelastic scenario, while the dashed line shows the spectrum for the best-fit point when $c_{\text{el}} = 0.8 c_{\text{in}}$.  LUX would see many events in its signal window; for models with a small elastic subcomponent, it could see as many as $\sim 40-50$ events in the winter (gray) and $\sim 70-80$ events in the summer (black).  

The inelastic and elastic scattering channels have remarkably similar nuclear recoil spectra because of the threshold behavior of the dominant inelastic channel and the form factor suppression of low momentum transfer events.  The shape of the spectra differ only at small recoil energies ($\lesssim$ 10 keVnr).  Because this is near the threshold energy of the experiment ($\sim 5$ keVnr), it is difficult to unambiguously distinguish the two contributions to the scattering rate with experiments such as LUX and XENON100.  In the following section, we show that additional information about the nuclear recoil direction is a critical component in understanding the dynamics of the dark sector.  


\section{Directional Detection}
\label{Sec: Directional Detection}

Directional detection experiments take advantage of the daily modulation in the direction of the dark matter wind in the lab frame, which arises from the Earth's rotation around the galactic center \cite{Spergel:1987kx, Ahlen:2009ev}.  In particular, the direction of the dark matter changes every twelve hours as the Earth rotates about its axis.  For the case of elastic scattering, the daily modulation amplitude can be nearly $\sim$100\%, compared to the $\sim$2.5\% change in the annual modulation amplitude.  

Measuring the strong angular dependence of the nuclear recoil can be an important tool for detecting dark matter \cite{Spergel:1987kx}.  Several directional detection experiments are currently running: DMTPC \cite{Sciolla:2009fb}, NewAge \cite{Miuchi:2007ga}, DRIFT \cite{Burgos:2008mv}, and MIMAC \cite{Santos:2007ga}.  To get reasonable angular resolution, the track left by the recoiling nucleus must be sufficiently long ($\sim 1$ mm).  This means that the detector material must be a gas, because liquid and crystalline detectors have scattering lengths that are too long.  All the current detectors are using  CF$_4$, except for DRIFT, which uses CS$_2$.  These detectors are specifically designed to look for elastic spin-dependent interactions, and so the use of atoms with high spin coupling, such as fluorine, is preferred.    

It was recently pointed out that directional detection experiments can serve as important tests of inelastic dark matter \cite{Finkbeiner:2009ug} if much heavier atoms are used in the detectors.  In this section, we show that such experiments are the key to distinguishing the contributions from elastic and inelastic scattering components in the dark matter sector.  To begin, we will briefly review how the rate equation derived in the previous section is generalized to include the angular dependence of the recoiling nucleus.  A more complete discussion of the theory can be found in \cite{Finkbeiner:2009ug, Gondolo:2002np}.  

Consider the lab frame where the incoming dark matter particle has a velocity $\vec{v}' = v \hat{x}$ and scatters off a nucleus at rest.  The recoiling nucleus has velocity $\vec{v}_R$ and makes an angle $\theta$ with the $\hat{x}$-axis. 
Energy and momentum conservation yield an expression for the recoil velocity,  
\begin{eqnarray}
v_R &=& \frac{2 \mu v}{m_N} \cos\theta, 
\label{eq: recoilvelocity}
\end{eqnarray}
which can be written in the frame-invariant form 
\begin{equation}
\hat{v} \cdot \hat{v}_R - \frac{v_{\text{min}}}{v} =0.
\end{equation}
The differential directional scattering rate per unit detector mass is
\begin{eqnarray}
\frac{d^2R}{dE_R d\cos\gamma} &=&\frac{\rho_0}{m_{\dm} m_N} \Big(v^2 \frac{d\sigma}{dE_R} \Big)  \\ 
&& \times  \int d^3v f(v) \delta(\vec{v}\cdot\hat{v}_R - \vec{v}_e\cdot \hat{v}_R- v_{\text{min}}).\nonumber
\label{eq: diffratedir}
\end{eqnarray}
\begin{figure}[b] 
   \centering
   \includegraphics[width=3.5in]{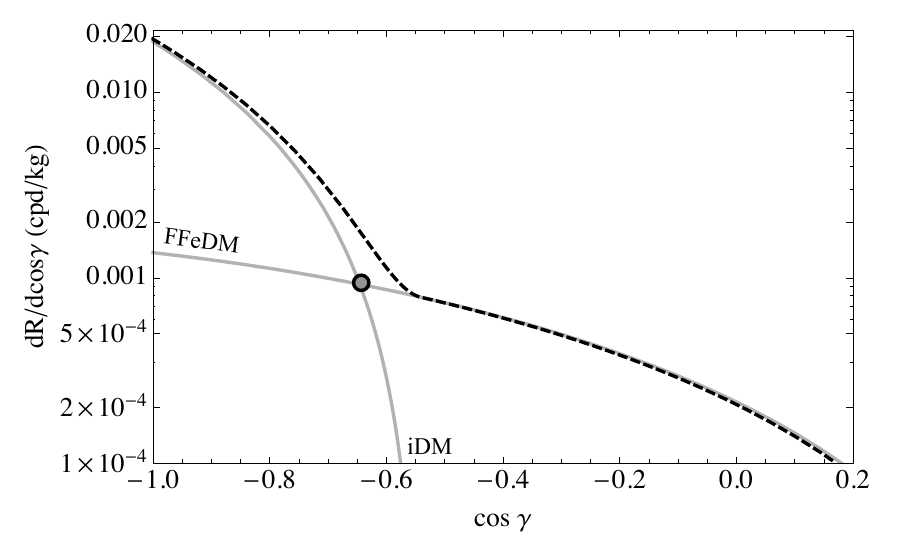} 
   \caption{$\cos\gamma$ spectrum during the summer for the best-fit parameters when $c_{\text{el}}  = 0.8 c_{\text{in}}$, assuming a detector of CF$_3$I.  The spectrum is taken for $E_r = 50$ keVnr.  The dashed line is the total from form factor-suppressed and inelastic scattering.  The shaded circle marks the ``cross-over" point where the elastic scattering starts to dominate over the inelastic scattering.}
   \label{Fig: Cos}
\end{figure}
The integral expression for the differential rate is an example of a Radon transform, the properties of which are reviewed in~\cite{Gondolo:2002np}.  The Radon transform for the modified halo distribution function we consider is 
\begin{equation}
\hat{f}(w) = \int\!\! d^3 v\; f(v) \delta( \vec{v}\cdot \hat{v}_R -w) = 2 \pi \int_w^{\infty}\!\!\!\! dv\; v f(v) ,
\end{equation}
where $w = \vec{v}_e\cdot \hat{v}_R + v_{\text{min}} = v_e \cos \gamma + v_{\text{min}}$.  Note that $\gamma$ is defined as the angle between the recoiling nucleus and the direction of the Earth's velocity $\vec{v}_e$.  Evaluating the integral for (\ref{eq: halo}), we find 
\begin{eqnarray}
\nonumber
\hat{f}(w)&=& \pi N \Bigg\{(w^2 - v_{\text{esc}}^2 ) e^{- (v_{\text{esc}}/v_0)^{2 \alpha}} \\
&&\hspace{-0.5in}+ \frac{v_0^2}{\alpha} \Bigg(\Gamma\Big[\frac{1}{\alpha}, \Big(\frac{w}{v_0}\Big)^{2 \alpha}\Big] - \Gamma\Big[\frac{1}{\alpha}, \Big(\frac{v_\text{esc}}{v_0}\Big)^{2 \alpha}\Big]\Bigg) \Bigg\} \Theta (v_{\text{esc}} - w).
\nonumber
\end{eqnarray}
There are several important features of this expression.  Firstly, the differential rate is peaked at $\cos \gamma = -1$.  this can also be seen explicitly from the rate equation, where the largest fraction of the parameter space is allowed when $\vec{v}_e\cdot \hat{v}_R = -1$ in the delta function, which makes sense because the rate is maximized when the Earth is moving in the same direction as the dark matter wind. 

The second important feature arises from the theta function in $\hat{f}(w)$.  In particular, there is a maximum value of $\cos \gamma$ above which the rate is zero:
\begin{equation}
\cos \gamma_{\text{max}} = \frac{v_{\text{esc}} -v_{\text{min}}}{v_e}.
\end{equation}
This cutoff depends on the minimum velocity, which is larger for inelastic scattering than elastic scattering.  As a result, the recoil spectrum for any inelastic scattering component will have a cut off \emph{below} that of any elastic scattering component.  This is illustrated in Fig.~\ref{Fig: Cos}, which shows the recoil spectrum for the best-fit parameters in a $c_{\text{el}}/c_{\text{in}} = 0.8$ theory at a recoil energy of 50 keVnr, which is near the threshold of most current experiments.  The rate for the inelastic interaction becomes negligible by $\cos\gamma \sim -0.5$.  There is a tail at larger values of $\cos \gamma$, which arises from the elastic scattering component.  

\begin{figure}[t] 
   \centering
   \includegraphics[width=3.5in]{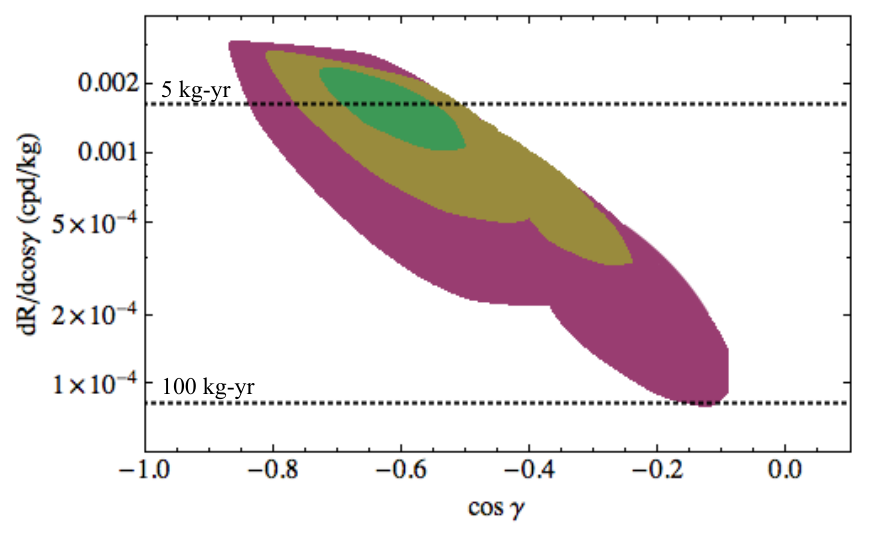} 
   \caption{The detection rate corresponding to the value of $\cos \gamma$ at the ``cross-over" point, where the elastic scattering starts to dominate over the inelastic scattering.  The shaded regions correspond to $c_{\text{el}}/c_{\text{in}}= $ 0.2 (magenta), 0.4 (yellow), and 0.8 (green).  The rate is calculated for summer at a recoil energy of $E_r = 50$ keVnr, assuming a detector of CF$_3$I.  The dashed lines show the projected sensitivity for the current DMTPC CF$_4$ detector after 5 and 100 kg-yr \cite{Sciolla:2008vp}. }
   \label{fig: CrossOver}
\end{figure}
Whether the elastic and inelastic scattering components can be distinguished depends on whether there are enough events that fall along the tail of the spectrum.  Fig.~\ref{fig: CrossOver} shows the expected rate at the value of $\cos \gamma$ where the transition from inelastic to elastic scattering occurs.  The spread arises from varying over the six unknowns of both the particle physics and halo profile models.  Even given the uncertainties of the halo profile distribution, one can obtain enough events on the $\cos\gamma$ tail for discovery with approximately 5-10 kg years-worth of data.  This is well within reach of current experiments; the dashed lines in Fig.~\ref{fig: CrossOver} show the projected sensitivity of the DMTPC experiment, assuming a CF$_4$ detector.  If a heavier detector material is used, as would be needed to test for inelastic dark matter, then the sensitivity should be even larger.

\section{Discussion}
\label{Sec: Discussion}

We have shown that directional detection experiments can distinguish mixed inelastic-elastic scattering scenarios that would otherwise be difficult to discover unambiguously, even with next-generation experiments such as LUX and XENON100.  In particular, directional detection experiments can detect elastic wide angle scatters that are kinematically forbidden in inelastic transitions. To evade current null experiments, a relatively large exposure of several $\text{kg-yrs}$ is needed, but this is well within reach of current detectors.  However, these experiments are currently optimized to look for spin-dependent elastic scattering and need to use heavier nuclei, such as iodine or xenon, to have sensitivity to inelastic recoils.  

Distinguishing different scattering mechanisms is critical for understanding the underlying symmetry structure of dark matter theories.  Many types of inelastic models can give small elastic scattering contributions.  For instance, nontrivial scattering mechanisms arise in models where the dark matter has a finite size, such as composite \cite{Alves:2009nf,Nussinov:1985xr, Khlopov:2008ki, Ryttov:2008xe}, atomic \cite{Kaplan:2009de}, mirror  \cite{Mohapatra:2001sx}, and quirky \cite{Kribs:2009fy} dark matter.  All these models have several scattering channels with a hierarchy of scales given by a series of higher dimensional operators.  Both elastic and inelastic operators may be allowed, and approximate discrete symmetries may cause the elastic scattering rate to be subdominant to the inelastic rate \cite{Lisanti:2009}.  

Another class of models that leads to form factor-suppressed elastic scattering is characterized by a pseudo-Goldstone mediator between the dark sector and the SM \cite{Chang:2009yt}.  In this case, additional dimension six operators in the Lagrangian may no longer be negligible in comparison to the standard spin-independent and spin-dependent operators.  These higher dimension operators lead to momentum suppression in the scattering rate.  There has been much interest recently in models with light mediators \cite{ArkaniHamed:2008qn, Morrissey:2009ur, Baumgart:2009tn, Cholis:2008qq, Finkbeiner:2009mi, Cheung:2009qd}, given the results from PAMELA \cite{Adriani:2008zr}, Fermi \cite{Abdo:2009zk}, ATIC \cite{:2008zzr}, and HESS \cite{Aharonian:2009ah}.  

If the dark sector does indeed have non-minimal structure, it may give rise to several types of scattering mechanisms.  Distinguishing these different scattering events is of fundamental importance for understanding the symmetry structure and unraveling the dynamics of the dark sector.  


\section*{Acknowledgements}
We thank Spencer Chang,  Liam Fitzpatrick,  Aaron Pierce,  Philip Schuster, Natalia Toro, and Neal Weiner for useful discussions.  We thank the Galileo Galilei Institute for their hospitality during the later stages of this work.
ML  and JGW are supported by the US DOE under contract number DE-AC02-76SF00515 and receive partial support from the Stanford Institute for Theoretical Physics.  ML is supported by an NSF fellowship.
JGW is partially supported by the US DOE's Outstanding Junior Investigator Award.  


\end{document}